\documentclass[aps,showpacs]{revtex4}
\usepackage{epsfig}
\newcommand{\ben}{\begin{eqnarray}}
\newcommand{\een}{\end{eqnarray}}
\newcommand{\nnu}{\nonumber\\}
\newcommand{\bef}{\begin{figure}[htb]\centering}
\newcommand{\eef}{\end{figure}}

\begin{document}
\title{Single transverse spin asymmetry of dilepton production near $Z^0$ pole}
\author{Zhong-Bo Kang}
\email{zkang@bnl.gov}
\affiliation{RIKEN BNL Research Center,
                 Brookhaven National Laboratory,
                 Upton, NY 11973, USA}
\author{Jian-Wei Qiu}
\email{jwq@iastate.edu}
\affiliation{Department of Physics and Astronomy,
                 Iowa State University,
                 Ames, IA 50011, USA}
                 
\begin{abstract}
Using the latest quark Sivers functions extracted from the global analysis of available data on single transverse spin asymmetry (SSA), we calculate the SSA of Drell-Yan inclusive production of lepton pairs of invariant mass $Q$ at both $4<Q<9$~GeV and $Q\sim M_Z$ in $p^\uparrow p$ collisions at RHIC energies.  We find that the features of the asymmetry $A_N$ for $Q\sim M_Z$ are significantly different from that when $4<Q<9$~GeV.  The $A_N$ near $Z^0$ pole is positive and sizable in the central rapidity region while the $A_N$ at low $Q$ is negative and only sizable in the forward rapidity region.  We show that the size of $A_N$ is sufficiently large for a wide range of $Q$ around the $Z^0$ pole, and even with the lepton pair's invariant mass integrated from $Q=70$~GeV to $110$~GeV, the $A_N$ is still close to 10 percent for the central and near forward rapidity region.  We argue that the SSAs of Drell-Yan inclusive dilepton production at low $Q$ and that near the $Z^0$ pole provide complementary information for testing the time-reversal modified universality (the sign change) of the Sivers functions.
\end{abstract}
\date{\today}
\maketitle

\section{Introduction}
Single transverse-spin asymmetries (SSAs) in both semi-inclusive deeply inelastic scattering (SIDIS) and high energy hadronic collisions have attracted much attention from both experimental and theoretical sides in recent years. From the parity and time-reversal invariance of the strong interaction dynamics, the measured asymmetries in these collisions should be directly connected to the transverse motion of partons inside a polarized hadron. Understanding the dynamics behind the measured asymmetries should have the profound impact on our knowledge of Quantum Chromodynamics (QCD) and hadron structure \cite{D'Alesio:2007jt}.

Two QCD-based approaches for analyzing the observed novel SSAs, the higher twist collinear factorization approach \cite{Efremov,qiu,koike,Qiu:2007ar} and the transverse momentum dependent (TMD) distribution approach \cite{Siv90,Collins93,Brodsky,MulTanBoe,TMD-dis,mulders}, have been
proposed and been applied extensively in phenomenological studies.  The first approach is based on the QCD collinear factorization \cite{CSS-fac,QS-fac}.  By generalizing the successful leading power QCD collinear factorization formalism to the next-to-leading power in the momentum transfer of the collision, non-vanishing SSAs could be generated by quantum interference between the real part and the imaginary part of the scattering amplitudes, and the SSA is proportional to  
twist-three transverse-spin-dependent multiparton correlation functions \cite{Efremov,qiu,Kang:2008ey}. This collinear factorization based approach is more relevant to the SSAs of the processes with all observed momentum transfers $Q\gg \Lambda_{\rm QCD}$. The other approach relies on the TMD factorization in QCD \cite{Collins:2002kn,JiMaYu04,CollinsMetz}.  In this approach, the non-vanishing SSAs are attributed to the spin-dependent TMD parton distribution functions (PDFs), known as the Sivers functions \cite{Siv90}, or the TMD fragmentation functions, known as the Collins functions \cite{Collins93}.  The TMD factorization based approach is more suitable for evaluating the SSAs of the processes with two very different momentum transfers, $Q_1\gg Q_2\gtrsim \Lambda_{\rm QCD}$.  The larger scale $Q_1$ is necessary for using perturbative QCD while the lower scale $Q_2$ makes the observable sensitive to the momentum of parton's transverse motion.  These two approaches have their own kinematic domains of validity, which are complementary to each others.  In the region where their validity overlaps, $Q_1\gg Q_2\gg \Lambda_{\rm QCD}$, these two seemingly different approaches should describe the same physics, and indeed, they were shown to be consistent for various processes where both factorizations were proved to be valid \cite{UnifySSA}. 

The predictive power of both approaches relies on the respective factorization.  
For the collinear factorization based approach, the predictive power is a consequence of the infrared safety of all-order perturbatively calculatable partonic hard parts and the universality of all factorized and nonperturbative PDFs and correlation functions \cite{CSS-fac,QS-fac}.  Unlike the collinearly factorized PDFs and correlation functions, the non-perturbative TMD PDFs or the Sivers functions are not exactly universal, instead, they have the time-reversal modified universality \cite{Collins:2002kn}.  It was shown from the parity and time-reversal invariance of QCD that the Sivers functions in SIDIS and that in Drell-Yan (DY) process should have the same functional form but an opposite sign - the time-reversal modified universality \cite{Collins:2002kn}.  From the measurement of Sivers functions in SIDIS, the sign of Drell-Yan SSA is uniquely predicted by the TMD factorization and the time-reversal modified universality of Sivers functions. 

The experimental test of the sign change of Sivers functions would provide a critical test for the TMD factorization approach and our understanding of the SSAs. Recently, HERMES and COMPASS experiments have performed measurements of Sivers functions in SIDIS \cite{HERMES,COMPASS}.  A new set of Sivers functions of various quark flavors was derived by the global analysis of available data \cite{Anselmino:2008sga}.  Future measurements of the SSAs in DY production have been planned \cite{Collins:2005rq,Anselmino:2009st}.  The SSAs of $W^\pm$ boson production in $p^\uparrow p$ collision at RHIC were also proposed to measure the DY-type Sivers functions \cite{ssa_W}.  Since it is difficult to reconstruct $W^\pm$ bosons by the current detectors at RHIC, we recently proposed to measure the SSA of single lepton decayed from $W$ bosons at RHIC to test the time-reversal modified universality of Sivers functions \cite{Kang:2009bp}.  We argued that with the large mass of $W$ bosons $M_W$ setting up the hard scale and the fact that the typical transverse momentum of $W$ bosons $|{\bf q}_\perp|$ produced at $\sqrt{s}=500$~GeV at RHIC is much less than the $W$ mass ($|{\bf q}_\perp|\ll M_W$), the single lepton production from $W$ boson decay at $\sqrt{s}=500$~GeV at RHIC is an ideal process for the TMD factorization.  We demonstrated that the SSA of single lepton decayed from $W$ bosons is sizable and its rapidity distribution is sensitive to the quark flavor of Sivers functions, and therefore, a good observable to test the sign change of the Sivers functions \cite{Kang:2009bp}.

In this paper, we extend our previous work on the SSA of single lepton from the decay of $W$ bosons to the SSA of inclusive lepton pair production with the pair's invariant mass near $Z^0$ boson mass $M_Z$ at $\sqrt{s}=500$~GeV at RHIC.  
In addition, we 
update the prediction of SSA of Drell-Yan lepton pair production with the pair's invariant mass $Q$ between the $J/\psi$ and $\Upsilon$ at RHIC energies by using the Sivers functions recently extracted from data of SIDIS experiments \cite{Anselmino:2008sga}.  We show that the asymmetry $A_N$ changes sign from low $Q$ to high $Q$ in mid-rapidity region and is sizable.  We find that the $A_N$ near $Z^0$ pole is significant for a wide range of $Q$ around $M_Z$, and the asymmetry could be used to test the time-reversal modified universality (the sign change) of the Sivers functions.

The rest of the paper is organized as follows.  In Sec.~\ref{theory}, we derive the theoretical formalism for calculating the SSA of inclusive lepton pair production in hadronic collisions in the TMD factorization approach.  In Sec.~\ref{plots}, we present our numerical predictions for the SSA of lepton pairs for RHIC kinematics.  Finally, we summarize our findings, corresponding conclusions and the potential improvements for our calculations in Sec.~\ref{summary}. 

\section{Single transverse spin asymmetry of dilepton production}
\label{theory}

The SSA of Drell-Yan lepton pair production via a virtual photon $\gamma^*$ has been studied previously in Refs.~\cite{Arnold:2008kf, Zhou:2009jm, Collins:2005rq, Anselmino:2009st}.  In this paper, we study the SSA of inclusive lepton pair production via both a virtual photon $\gamma^*$, a $Z^0$ boson, and their interference, 
\ben
A^\uparrow(p_A, {\bf S})+B(p_B)\to \left[\gamma^*, Z\to \right]\ell^+\ell^-(Q)+X\, ,
\label{process}
\een
in hadronic collisions between a polarized proton $A$ of momentum $p_A$ and transverse-spin vector ${\bf S}$ and an unpolarized proton $B$ of momentum $p_B$.  The scale of hard collision $Q$ in Eq.~(\ref{process}) is the invariant mass of the observed lepton pair.  For testing the TMD factorization and the time-reversal modified universality of Sivers functions, we concentrate on the kinematic region where the transverse momentum of the produced lepton pair $|{\bf q}_\perp|$ is much less than the mass of the pair ($|{\bf q}_\perp|\ll Q$), which is the region of phase space where the TMD factorization formalism is expected to be valid \cite{JiMaYu04}. 

In terms of the TMD factorization, the leading order cross section for dilepton production can be written as 
\ben
\frac{d\sigma_{A^\uparrow B\to \ell^+\ell^-+X}({\bf S})}{dQ^2dy\, d^2{\bf q_\perp}}=\sum_q\int d^2{\bf k}_{a\perp}d^2{\bf k}_{b\perp}\delta^2({\bf q}_\perp-{\bf k}_{a\perp}-{\bf k}_{b\perp})f_{q/A^\uparrow}(x_a,{\bf k}_{a\perp},{\bf S})f_{\bar{q}/B}(x_b,k_{b\perp})\, \hat{\sigma}_0^{q\bar{q}},
\label{main}
\een
where $\sum_q$ runs over all light (anti)quark flavors, $y$ is the rapidity of the observed lepton pair, and $k_{b\perp}=|{\bf k}_{b\perp}|$ is the magnitude of parton transverse momentum.  The lowest order partonic cross section $\hat{\sigma}_0^{q\bar{q}}$ in Eq.~(\ref{main}) receives the contribution from both the virtual photon and the $Z^0$ boson, as well as their interference, and is given by,
\ben
\hat{\sigma}_0^{q\bar{q}}=\frac{1}{N_c s}\left[e_q^2\frac{4\pi\alpha^2}{3Q^2}
+e_q\frac{4\alpha}{3}\frac{G_F}{\sqrt{2}}v_qv_\ell{\mathcal Re}(R)
+\frac{Q^2}{3\pi}\left(\frac{G_F}{\sqrt{2}}\right)^2\left(v_q^2+a_q^2\right)\left(v_\ell^2+a_\ell^2\right)|R|^2
\right],
\label{sigma0}
\een
where $s=(p_A+p_B)^2$, $\alpha$ and $G_F$ are the electro-magnetic and Fermi weak coupling constants, respectively, $e_q$ is the quark fractional charge of flavor $q$, $N_c=3$ is the number of color, $v_{q(\ell)}$ and $a_{q(\ell)}$ are the vector and axial couplings of the $Z^0$ boson to the quark (lepton), respectively, and 
\ben
R=\frac{M_Z^2}{Q^2-M_Z^2+i\,M_Z\Gamma_Z},
\een
with $M_Z$ and $\Gamma_Z$ the mass and decay width of the $Z^0$ boson, respectively. The parton momentum fractions in Eq.~(\ref{main}) are given by,
\ben
x_a=\frac{Q}{\sqrt{s}}e^{y},\qquad
x_b=\frac{Q}{\sqrt{s}}e^{-y}
\een
to the leading power in $q_\perp^2/Q^2$.

Following the notation of Ref.~\cite{Anselmino:2008sga}, we expand the spin dependent TMD quark distribution as,
\ben
f_{q/A^\uparrow}(x,{\bf k}_\perp,{\bf S})
&\equiv &f_{q/A}(x,k_\perp) 
+
\frac{1}{2}\Delta^N f_{q/A^\uparrow}(x,k_\perp)\,
{\bf S}\cdot (\hat{p}_A\times \hat{\bf k}_\perp),
\label{sivers}
\een
where $k_\perp=|{\bf k}_\perp|$, $\hat{p}_A$ and $\hat{\bf k}_\perp$ are the unit vectors of $p_A$ and ${\bf k}_\perp$, respectively, $f_{q/A}(x,k_\perp)$ is the spin-averaged TMD quark distribution function of flavor $q$, and $\Delta^N f_{q/A^\uparrow}(x,k_\perp)$ is the Sivers function of flavor $q$.  Substituting Eq.~(\ref{sivers}) to Eq.~(\ref{main}), we obtain for the spin-dependent cross section as
\ben
\frac{d\Delta\sigma_{A^\uparrow B\to \ell^+\ell^-+X}({\bf S})}
     {dQ^2dy\, d^2{\bf q_\perp}}
&\equiv&\frac{1}{2}
\left[
\frac{d\sigma_{A^\uparrow B\to \ell^+\ell^-+X}({\bf S})}
{dQ^2dy\, d^2{\bf q_\perp}}
-
\frac{d\sigma_{A^\uparrow B\to \ell^+\ell^-+X}({\bf -S})}
{dQ^2dy\, d^2{\bf q_\perp}}
\right]
\nnu
&=&
\frac{1}{2}\sum_q\int d^2{\bf k}_{a\perp}d^2{\bf k}_{b\perp}\delta^2({\bf q}_\perp-{\bf k}_{a\perp}-{\bf k}_{b\perp})\,
{\bf S}\cdot (\hat{p}_A\times \hat{\bf k}_{a\perp})
\nnu
&\ & \times
\Delta^Nf_{q/A^\uparrow}(x_a,k_{a\perp})\, f_{\bar{q}/B}(x_b,k_{b\perp})\,\hat{\sigma}_0^{q\bar{q}}\, ,
\label{dep}
\een
and corresponding spin-averaged cross section as
\ben
\frac{d\sigma_{AB\to \ell^+\ell^-+X}}{dQ^2dy\, d^2{\bf q_\perp}}
&=&
\sum_q\int d^2{\bf k}_{a\perp}d^2{\bf k}_{b\perp}\delta^2({\bf q}_\perp-{\bf k}_{a\perp}-{\bf k}_{b\perp})\,
f_{q/A}(x_a,k_{a\perp})\, f_{\bar{q}/B}(x_b,k_{b\perp})\,
\hat{\sigma}_0^{q\bar{q}} \, .
\label{avg}
\een
From Eqs.~(\ref{dep}) and (\ref{avg}), we obtain the conventionally defined SSA as
\ben
A_N=
\frac{d\Delta\sigma_{A^\uparrow B\to \ell^+\ell^-+X}({\bf S})}{dQ^2dy\, d^2{\bf q_\perp}}
\left/
\frac{d\sigma_{AB\to \ell^+\ell^-+X}}{dQ^2dy\, d^2{\bf q_\perp}}.\right.
\label{AN}
\een
The commonly used weighted asymmetry is defined as
\ben
A_N^{\sin(\phi-\phi_s)}= 2
\int_0^{2\pi}d\phi\, \sin(\phi-\phi_s)\,
\frac{d\Delta\sigma_{A^\uparrow B\to \ell^+\ell^-+X}({\bf S})}{dQ^2dy\, d^2{\bf q_\perp}}
\left/
\int_0^{2\pi}d\phi \frac{d\sigma_{AB\to \ell^+\ell^-+X}}{dQ^2dy\, d^2{\bf q_\perp}},
\right.
\label{weight}
\een
where $\phi$ is the azimuthal angle of the lepton pair of transverse momentum ${\bf q}_\perp$ and $\phi_s$ is the azimuthal angle of the spin vector ${\bf S}$ of incoming polarized hadron.

In order to evaluate the SSAs in Eqs.~(\ref{AN}) and (\ref{weight}) for inclusive lepton pair production in $p^\uparrow p$ collisions, we adopt the parameterization of TMD parton distributions introduced in Ref.~\cite{Anselmino:2008sga},
\ben
f_{q/h}(x,k_\perp) 
&=& f_q(x)\,
\frac{1}{\pi\langle k_\perp^2\rangle}\, 
e^{-k_\perp^2/\langle k_\perp^2\rangle} ,
\label{TMDavg} \\
\Delta^N f_{q/h^\uparrow}^{\rm SIDIS}(x,k_\perp)
&=&
2\,{\cal N}_q(x)\,h(k_\perp)\,
f_{q/h}(x,k_\perp) ,
\label{Sivers_f} \\
h(k_\perp)
&=&
\sqrt{2e}\, \frac{k_\perp}{M_1}\, 
e^{-k_\perp^2/M_1}
\een
where $f_q(x)$ is the standard unpolarized parton distribution 
of flavor $q$, 
$\langle k_\perp^2\rangle$ and $M_1$ are fitting parameters, and
${\cal N}_q(x)$ is a fitted distribution given in 
Ref.~\cite{Anselmino:2008sga}. 

With the Gaussian ansatz for the transverse momentum dependence of the TMD distributions, one could carry out
the integration $d^2{\bf k}_{a\perp}d^2{\bf k}_{b\perp}$ in Eqs.~(\ref{dep}) and (\ref{avg}) analytically, and obtain
\ben
A_N={\bf S}\cdot (\hat{p}_A\times {\bf q}_\perp)\frac{\sqrt{2e}}{M_1}
\frac{2\langle k_s^2\rangle^2}
     {[\langle k_\perp^2\rangle + \langle k_s^2\rangle]^2}\,
e^{-\left[
     \frac{\langle k_\perp^2\rangle-\langle k_s^2\rangle}
          {\langle k_\perp^2\rangle + \langle k_s^2\rangle}
    \right] 
    \frac{\mathbf{q}_\perp^2}{2\langle k_\perp^2\rangle}}
\frac{\sum_q \hat{\sigma}_0^{q\bar{q}}\left[-{\cal N}_q(x_a)\right]f_{q/A}(x_a)f_{\bar{q}/B}(x_b)}
    {\sum_q \hat{\sigma}_0^{q\bar{q}}f_{q/A}(x_a)f_{\bar{q}/B}(x_b)},
\label{ANgauss}
\een
where $\langle k_s^2\rangle=M_1^2\, \langle k_\perp^2\rangle/[M_1^2+\langle k_\perp^2\rangle]$ and the ``$-$'' sign in front of ${\cal N}_q(x_a)$ is due to the sign difference between the Sivers function in DY process and that in SIDIS process. 

For our numerical results in next section, we choose a frame in which the polarized hadron 
$A$ moves in the $+z$-direction, and ${\bf S}$ and ${\bf q}_\perp$ along $y$- and $x$-directions, respectively.  It is important to realize that the weighted asymmetry defined in Eq.~(\ref{weight}) differ from $A_N$ in Eq.~(\ref{AN}) by a minus sign, 
\ben
A_N^{\sin(\phi-\phi_s)}=-A_N 
\een
in our choice of frame. 

\section{Phenomenology}
\label{plots}

In this section, we present our numerical estimates for the SSA of inclusive lepton pair production in $p^\uparrow p$ collisions at RHIC energies by using Eq.~(\ref{ANgauss}).   We use GRV98LO parton distribution \cite{GRV98} to be consistent with the usage of the TMD distributions of Ref.~\cite{Anselmino:2008sga}. 

In Fig.~\ref{lowQ}, we plot the asymmetry $A_N$ for the lepton pair production with the pair's transverse momentum integrated up to $q_T\equiv|{\bf q}_\perp| = 1$~GeV at RHIC energy $\sqrt{s}=200$~GeV (solid lines) and $\sqrt{s}=500$~GeV (dashed lines).  In the figure on the left, we plot $A_N$ as a function of rapidity $y$ of the lepton pair with the pair's invariant mass $Q$ integrated from 4~GeV to 9~GeV.  Similarly, in the figure on the right, we plot 
$A_N$ as a function of the pair's invariant $Q$ with the pair's rapidity integrated from $0$ to $3$.  Since the mass of $Z^0$ boson is so much larger than the lepton pair's invariant mass $Q$ considered in Fig.~\ref{lowQ}, the $A_N$ at $\sqrt{s}=200$~GeV (solid lines in two figures) are effectively the same as those presented in Ref.~\cite{Anselmino:2009st}, in which only the virtual photon channel was considered.  That is, as expected, the $Z^0$ boson's contribution to the production rate as well as the asymmetry is not important for the region where $Q\ll M_Z$. 
\bef
\psfig{file=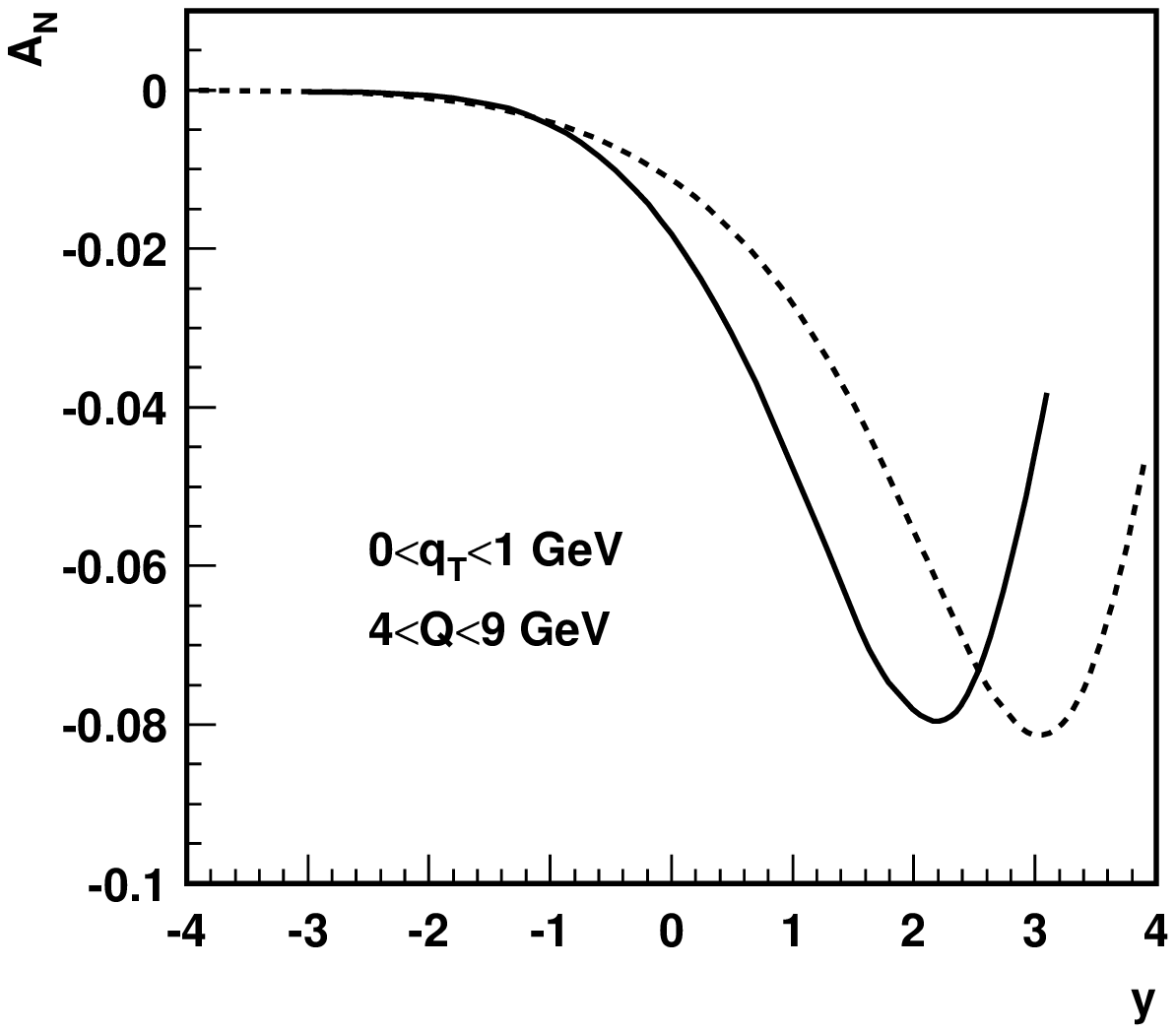,width=2.5in}\hskip 0.2in
\psfig{file=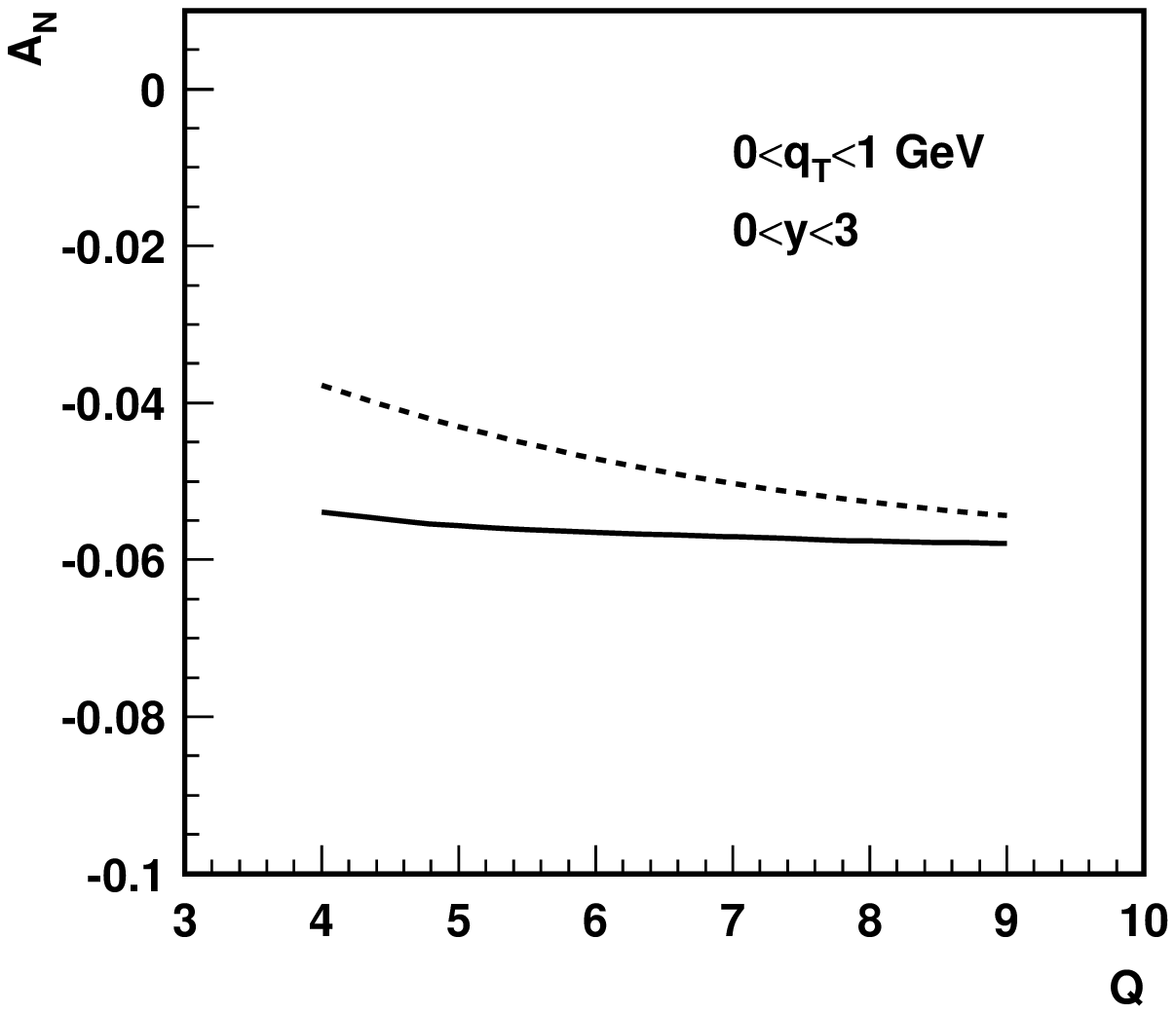,width=2.5in}
\caption{SSA of lepton pair production as a function of the pair's rapidity $y$ (left) and invariant mass $Q$ (right) at $\sqrt{s}=200$ GeV (solid lines) and $\sqrt{s}=500$ GeV (dashed lines). We have integrated the pair's transverse momentum from 0 to 1 GeV.}
\label{lowQ}
\eef

As shown in Fig.~\ref{lowQ}, $A_N$ of inclusive lepton pair production in this small $Q$ region is negative. Unlike the $W^\pm$-boson production (or the lepton production from the decay of $W$-boson) \cite{Kang:2009bp}, Drell-Yan inclusive lepton pair production via a virtual photon or a $Z^0$ boson is not very good in separating contributions from different quark or antiquark flavors.  The SSA in this low $Q$ region is dominated by the virtual photon channel and is proportional to a {\it sum} of quark (antiquark) Sivers function of {\it all} flavors weighted by the quark fractional charge square $e_q^2$ and corresponding spin-averaged antiquark (quark) distribution.  Since the $u$-quark and $d$-quark Sivers functions have the opposite sign \cite{Anselmino:2008sga}, the sign and size of the SSA of inclusive dilepton production is a result of an incomplete cancelation between contributions from different quark flavors.  For the latest parameterization of the Sivers functions obtained in Ref.~\cite{Anselmino:2008sga}, the $u$-quark Sivers function for Drell-Yan type processes is negative  while $d$-quark Sivers function is positive, and the absolute size of $u$-quark Sivers function is slightly smaller than $d$-quark Sivers function.  Since the contribution to the SSA from $u$-quark Sivers function is weighted by a larger fractional charge square, it wins over the contribution from $d$-quark Sivers function, and consequently, we have a negative $A_N$ as shown in Fig.~\ref{lowQ}.  The rapidity dependence of $A_N$ is a consequence of the $x$-dependence of the Sivers functions extracted from the global fitting in Ref.~\cite{Anselmino:2008sga}.  The measurement of the rapidity dependence of the SSA of Drell-Yan lepton pair production could provide information on both the sign and the functional form of the Sivers functions to test the time-reversal modified universality of the Sivers functions.

In Fig.~\ref{lowQ}, we also plot the SSA of Drell-Yan lepton pair production at $\sqrt{s}=500$~GeV (dashed lines).  It is clear that the asymmetries at 
$\sqrt{s}=500$~GeV carry the same features as those at $\sqrt{s}=200$~GeV.  However, due to the larger collision energy while keeping the same range of $q_T$ and $Q$, the same rapidity $y$ corresponds to different values of  momentum fractions of colliding partons. That is, the rapidity dependence of the SSA at different collision energy provides additional information on the functional form of the Sivers functions.

The SSA of Drell-Yan lepton pair production is small in the central region of the collision ($y\sim 0$) due to the cancelation of the contributions from different quark flavors.  The asymmetry becomes sizable for the experimental measurement only in the forward region ($y\sim 1-3$), as shown in Fig.~\ref{lowQ}, where the production rate for the lepton pair is significantly reduced.  Since the hard scale $Q$ is not very large in this low $Q$ region, the theoretical uncertainty for (or correction to) the leading power TMD factorization formalism becomes larger in the large $y$ forward region.  In principle, the TMD factorization works better when the hard scale $Q$ is much larger than $q_T$.  But, in practice, the rate for lepton pair production quickly diminishes when the invariant mass $Q$ and/or rapidity $y$ increase while the collision energy $\sqrt{s}$ is fixed.  However, with the new collision energy at $\sqrt{s}=500$~GeV, RHIC can produce sufficient numbers of $W^\pm$ and $Z^0$
bosons.  The large mass of $W^\pm$ and $Z^0$ makes the corrections to the TMD factorization formalism smaller and the theoretical calculations more reliable.  The SSA of inclusive lepton pair production at $Q=M_Z$, which is mainly from the decay of $Z^0$, could be a very good complementary observable to the SSA of $W^\pm$ production to test the TMD factorization in QCD and the time-reversal modified universality of the Sivers functions. 

In Fig.~\ref{Zpole}, we plot the SSA of lepton pair production at $Z^0$ pole ($Q=M_Z$) at RHIC at $\sqrt{s}=500$ GeV as a function of the pair's transverse momentum $q_T$ (left) and the rapidity $y$ (right).  We find that the SSA of inclusive lepton pair production at $Z^0$ pole is positive, in contrast to the negative SSA in Fig.~\ref{lowQ}.  Different from the low $Q$ region, the production of lepton pair at $Q=M_Z$ is dominated by the $Z^0$ channel.  Although both $u$- and $d$-quark Sivers function in the $Z^0$ channel contribute to the SSA in the same way as that in the virtual photon channel, the relative weight between their contributions is different.  Instead of the fractional charge square $e_q^2$ in the virtual photon channel, the contribution of quark flavor $q$ in the $Z^0$ channel is weighted by $v_q^2+a_q^2$ with
\ben
v_u&=&\frac{1}{2}-\frac{4}{3}\sin^2\theta_W
\qquad
a_u=\frac{1}{2},
\nnu
v_d&=&-\frac{1}{2}+\frac{2}{3}\sin^2\theta_W
\qquad
a_d=-\frac{1}{2}.
\een
Unlike the virtual photon channel, where $e_u^2 > e_d^2$, the relative weight for the contribution of $u$- and $d$-quark Sivers function in the $Z^0$ channel is opposite, $v_u^2+a_u^2 < v_d^2+a_d^2$.  Since the absolute value of $d$-quark Sivers function is larger than that of $u$-quark Sivers function as found in Ref.~\cite{Anselmino:2008sga}, the contribution from the $d$-quark Sivers function wins over that from $u$-quark Sivers function in the $Z^0$ channel,   and consequently, $A_N > 0$ when the lepton pair's invariant mass is near the  $Z^0$ pole, $Q\sim M_Z$, as shown in Fig.~\ref{Zpole}.
\bef
\psfig{file=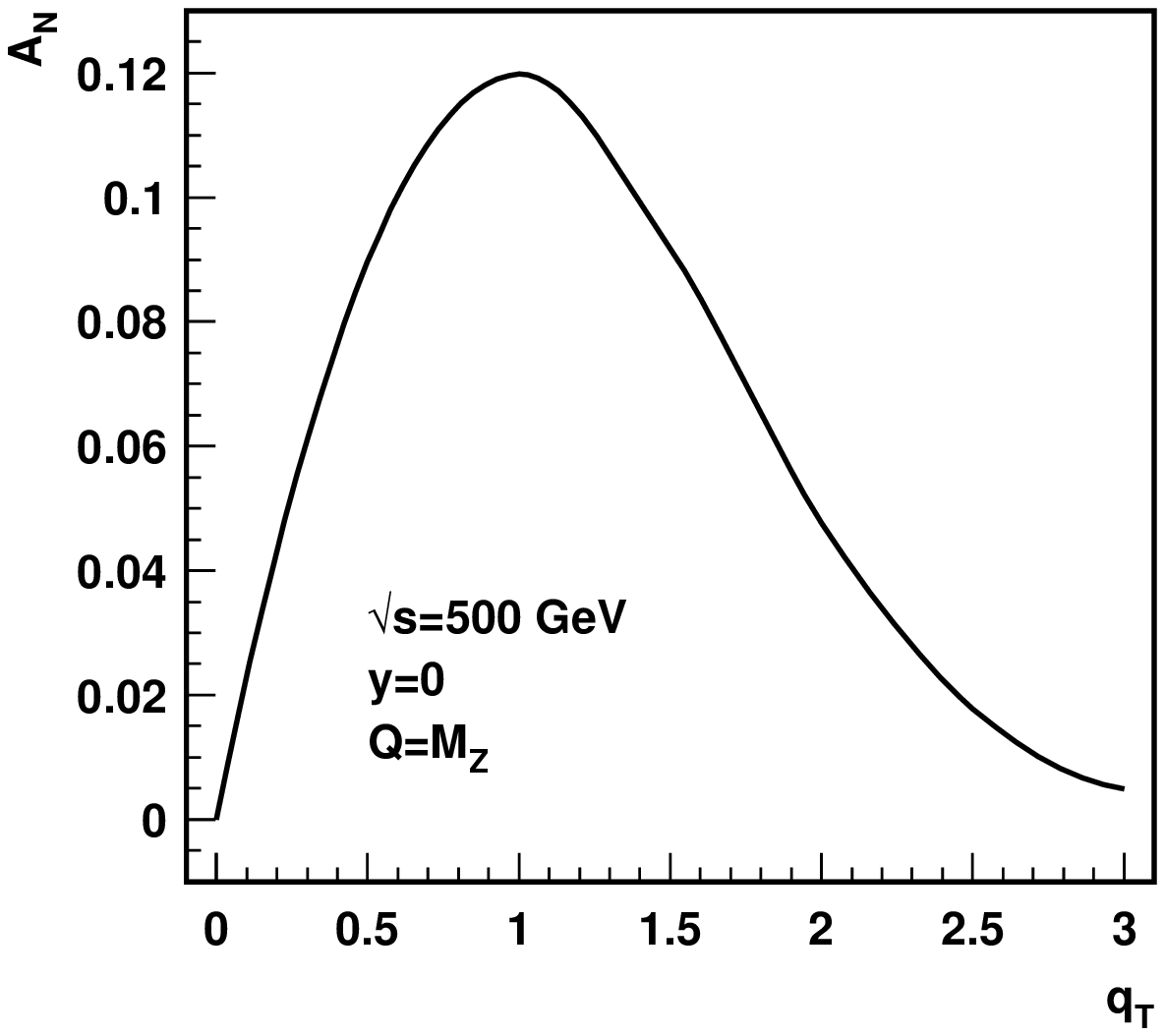,width=2.5in}\hskip 0.2in
\psfig{file=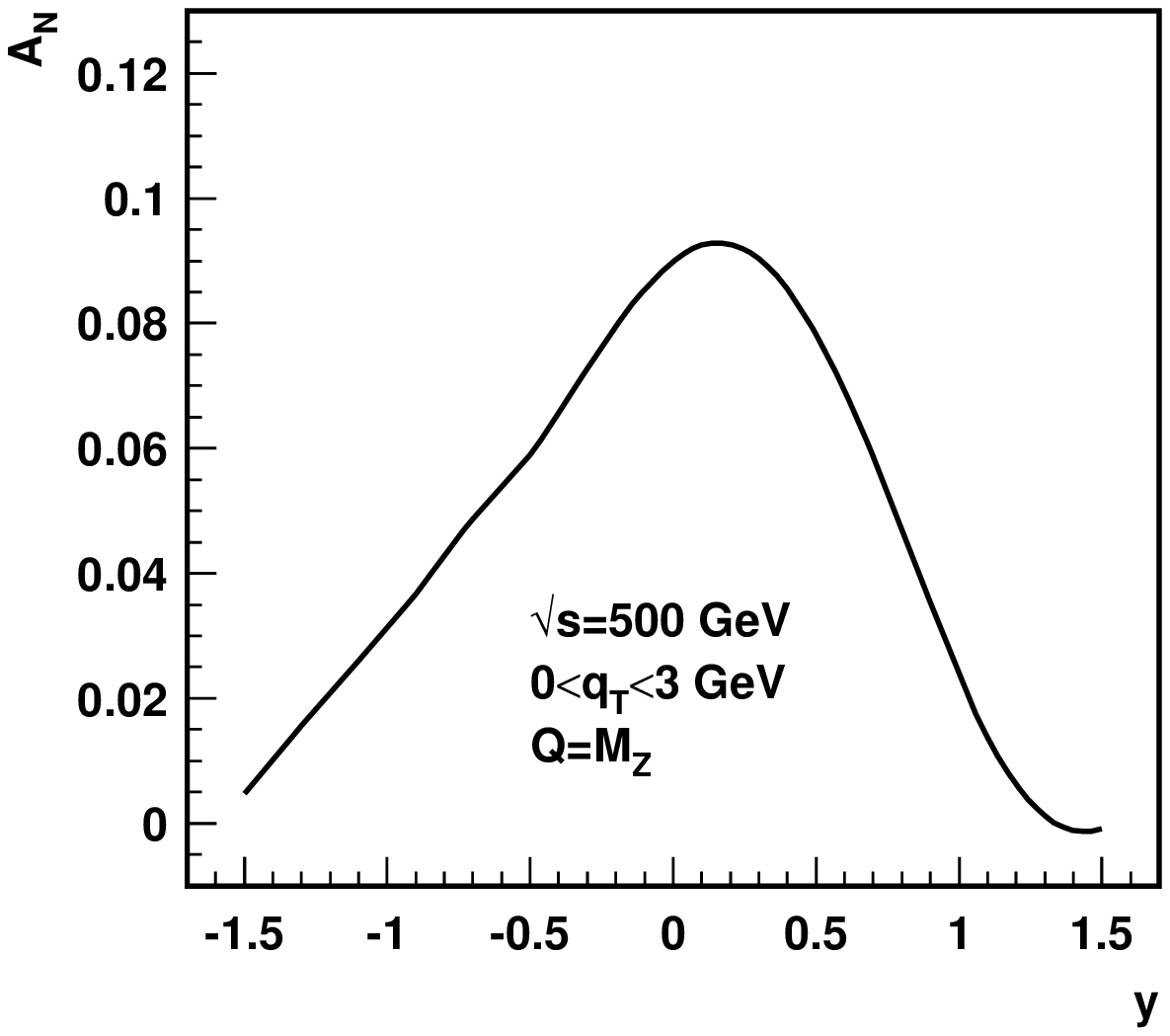,width=2.5in}
\caption{SSA of lepton pair production at $Z^0$ pole $Q=M_Z$ as a function of the pair's transverse momentum $q_T$ (left) and rapidity $y$ at $\sqrt{s}=500$ GeV.}
\label{Zpole}
\eef

As shown in Fig.~\ref{Zpole}, $A_N$ of inclusive lepton pair production with the pair's invariant mass $Q=M_Z$ in $p^\uparrow p$ collisions at $\sqrt{s}=500$~GeV is very significant at $q_T\sim 1$~GeV in the central $y$ region.  The absolute size of the asymmetry in the central region is much larger than those shown in Fig.~\ref{lowQ} for the low $Q$ region.  The SSA of inclusive lepton pair production near $Z^0$ pole could be an equally good observable as the low $Q$ Drell-Yan lepton pair production to test the time-reversal modified universality of the Sivers functions, if there is a sufficient rate of producing $Z^0$ bosons at RHIC.

To better understand the sign change of $A_N$ of inclusive lepton pair production from the low $Q$ region to the region near $Z^0$ pole, we plot the $A_N$ as a function of $Q$ in Fig.~\ref{tran} (left).  The figure clearly demonstrates the transition from the negative $A_N$ at a low invariant mass $Q$ to the positive $A_N$ when $Q$ is near $Z^0$ pole.  In our calculation, the lowest order partonic cross section, $\hat{\sigma}_0^{q\bar{q}}$ in Eq.~(\ref{sigma0}), determines the relative weight between the contributions of $u$-quark and $d$-quark Sivers function at a given dilepton invariant mass $Q$.
\bef
\psfig{file=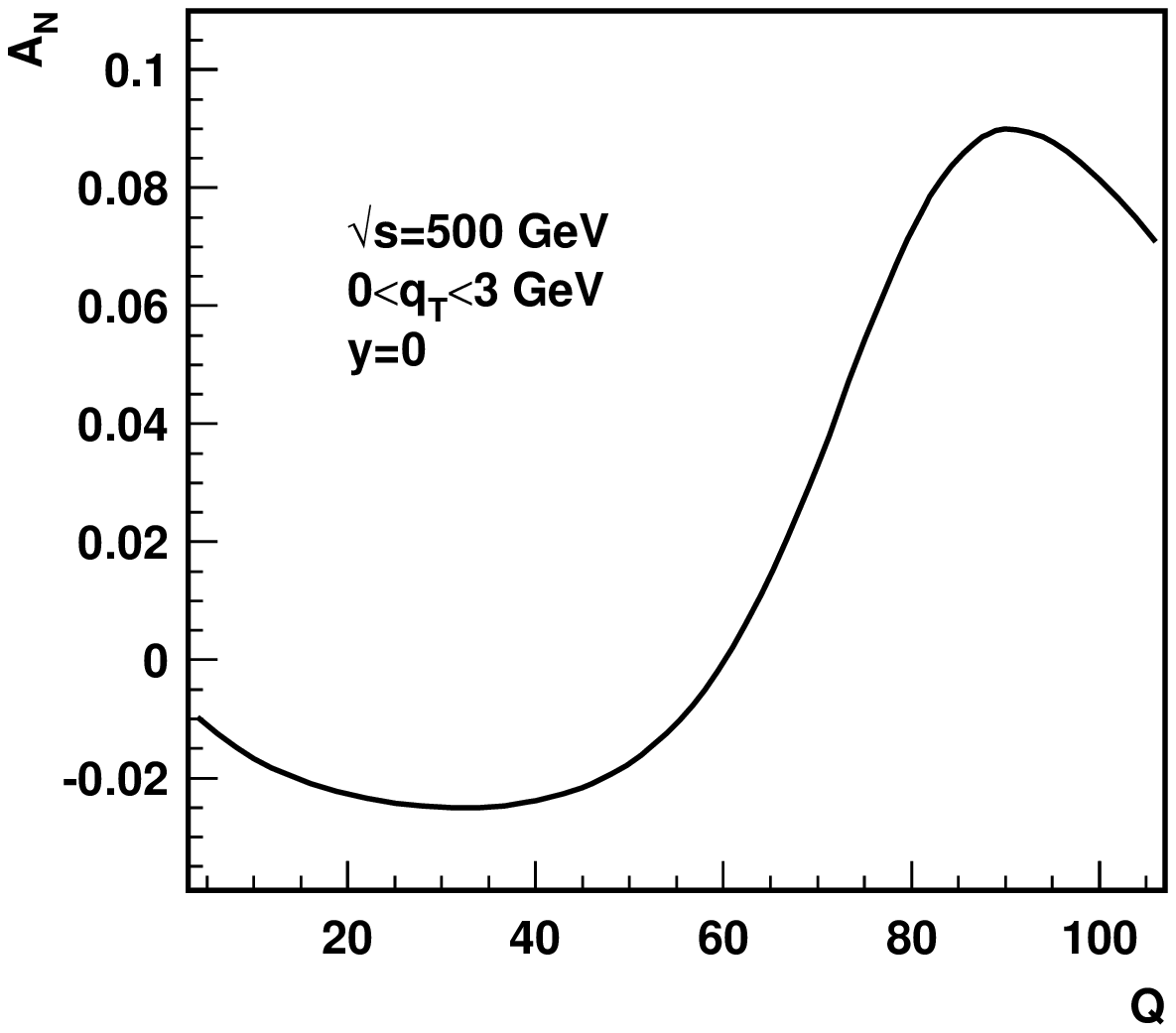,width=2.5in}\hskip 0.2in
\psfig{file=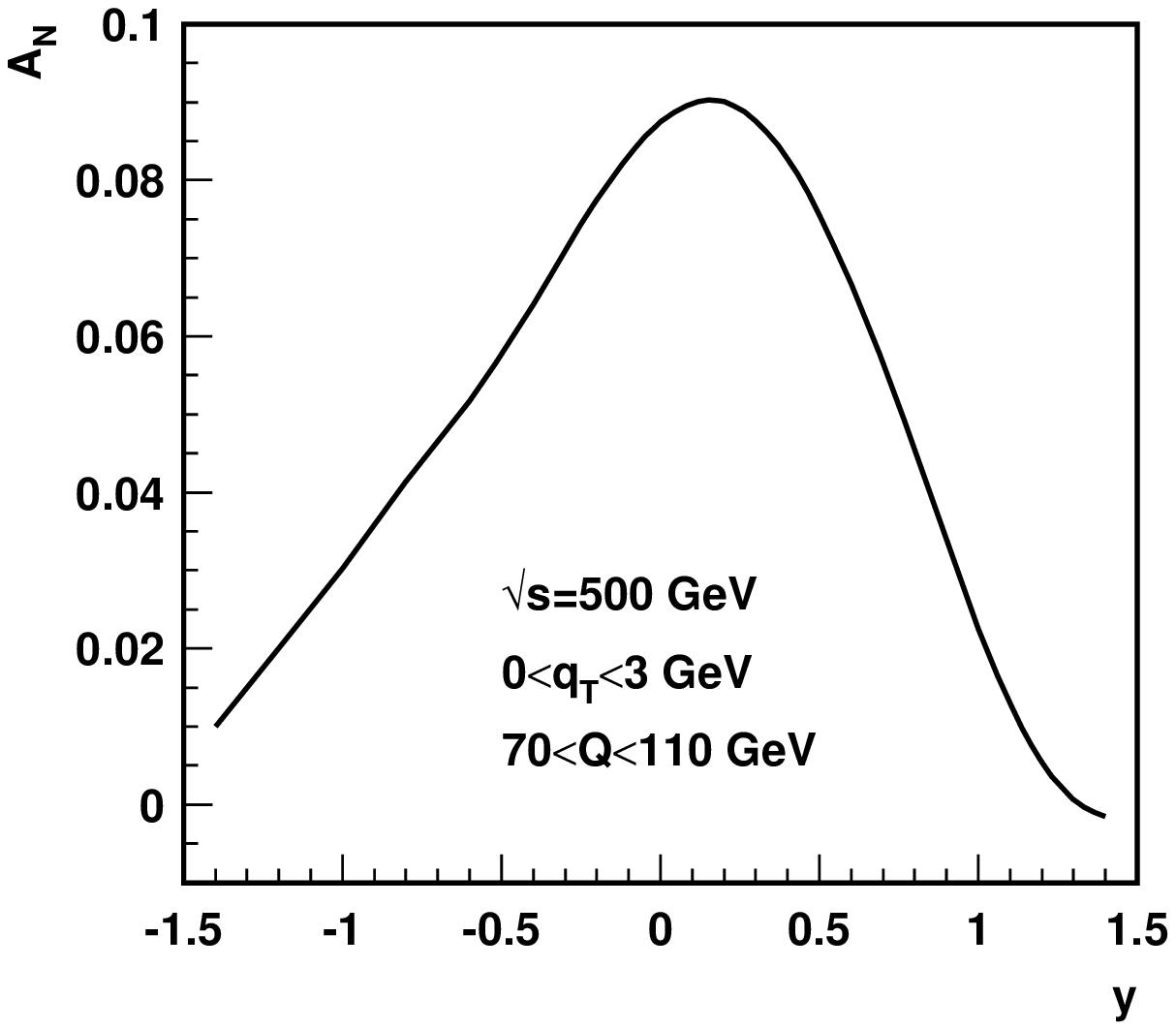,width=2.5in}
\caption{Left: SSA of lepton pair production as a function of the pair's invariant mass 
$Q$. Right: SSA of lepton pair accumulated around $Z^0$ pole as a function of rapidity $y$.}
\label{tran}
\eef

From Fig.~\ref{tran} (left), the $A_N$ is sufficiently large even when the pair's invariant mass is not exactly sitting at the $Z^0$ pole.  Therefore, we could have a relatively wide bin of $Q$ to improve the event rate of inclusive lepton pair production at large $Q$.  In Fig.~\ref{tran} (right), we plot the SSA as a function of rapidity $y$ for all lepton pairs whose invariant mass $Q$ are between 70 GeV to 110 GeV.  It is clear that the SSA of inclusive lepton pair production at $\sqrt{s}=500$~GeV for a wide bin of invariant mass $Q$ around the $Z^0$ pole is close to 10\% at the central rapidity, and could be measurable in the future at RHIC.

\section{Conclusion}
\label{summary}

We studied, in terms of TMD factorization formalism, the single transverse spin asymmetry of inclusive lepton pair production in $p^\uparrow p$ collisions at RHIC energies for the kinematic region where the lepton pair's transverse momentum is much less than the pair's invariant mass, $q_T\ll Q$, the condition that is required for justifying the TMD factorization.  In our calculation of the SSA, we included contributions to the lepton pair production from both the virtual photon and $Z^0$ boson channel as well as the contribution from their interference.  

By using the latest Sivers functions extracted from the global analysis of available data \cite{Anselmino:2008sga}, we evaluated the SSA of Drell-Yan lepton pair production for $4<Q<9$~GeV at $\sqrt{s}=200$~GeV and 500~GeV.  The asymmetry $A_N$ at this low $Q$ region is close to zero for the central and negative rapidity region due to the cancelation between the contribution from $u$-quark and $d$-quark Sivers functions.  The asymmetry $A_N$ is negative and sizable in the forward rapidity region ($y\sim 1-3$) \cite{Anselmino:2009st}.  

We also calculated the SSA for inclusive production of lepton pairs with the pairs' invariant mass near $Z^0$ boson pole in $p^\uparrow p$ collisions at $\sqrt{s}=500$~GeV.  We found that the features of the asymmetry $A_N$ for $Q\sim M_Z$ are significantly different from that when $4<Q<9$~GeV.  As shown in Figs.~\ref{lowQ} and \ref{Zpole}, the asymmetry $A_N$ near $Z^0$ pole is positive and sizable in the central rapidity region while the $A_N$ at low $Q$ is negative and only sizable in the forward rapidity region.  The sign change of the asymmetry is due to the transition of contributions from the virtual photon channel dominance to $Z^0$ channel dominance when $Q$ increases, which leads to the change of the relative weight of the contributions from $u$-quark and $d$-quark Sivers function that have opposite sign.   Since the effective weight of the different flavor contributions to the asymmetry $A_N$ varies when the hard scale $Q$ changes, we conclude that the SSAs at low $Q$ region and that near $Z^0$ pole provide complementary information on both the sign and the functional form of the Sivers functions as well as their flavor dependence.

We noticed that the size of asymmetry $A_N$ for inclusive lepton pair production is sufficiently large for a wide range of $Q$ around the $Z^0$ pole.  We found that the $A_N$ with lepton pair's invariant mass integrated from $Q=70$~GeV to $110$~GeV is close to 10 percent for the central and near forward rapidity region.  This large bin width for the lepton pair's invariant mass helps to increase the event rate, and the size of $A_N$ should be measurable in the future at RHIC.  The successful measurement of SSA of inclusive lepton pair production could provide a crucial test of the time-reversal modified universality (or the sign change) of Sivers functions and the validity of TMD factorization approach to the novel phenomenon of SSA.

Our results and conclusions for the SSA of inclusive lepton pair production rely on the TMD factorization and the accuracy of the extracted spin-averaged TMD parton distributions and the Sivers functions as well as their factorization scale dependence.  Unlike the parton distribution functions in the collinear factorization approach, whose factorization scale dependence is well-understood and determined by the DGLAP evolution equation, the factorization scale dependence of TMD parton distributions is not yet fully understood.  Without reliable theoretical input, authors of Ref.~\cite{Anselmino:2008sga} assumed that the Sivers functions obey the spin-averaged DGLAP evolution equation in their global analysis of extracting TMD parton distributions.  Although it is difficult to judge how good the assumption is, we noticed that the evolution equation for the twist-3 quark-gluon correlation functions relevant to the SSA in collinear factorization approach, which are one-to-one related to the first moment of the quark Sivers functions \cite{Boer:2003cm}, is different from the spin-averaged DGLAP evolution equation \cite{Kang:2009bp,T3evolution}.  But, the difference is relatively small as long as the parton momentum fraction $x$ is not too small \cite{Kang:2009bp}.   Therefore, we think that using the Sivers functions from Ref.~\cite{Anselmino:2008sga}, which fit all existing data, is a reasonable choice that we can make now, and the general features of the SSA of inclusive lepton pair production presented in this paper should be valid.  For the precision test of TMD factorization in QCD, it is of course very important to understand the factorization scale dependence of the Sivers functions or TMD parton distributions in general, which is beyond the scope of this paper. 

When the hard scale of the collision, $Q$, the invariant mass of the produced lepton pair is so much larger than the observed transverse momentum of the pair, $Q\gg q_T$, the $Q$ and $q_T$ dependence of the production rate is dominated by the gluon shower right before the hard quark-antiquark annihilation to produce the lepton pair. The $Q$ and $q_T$ dependence of the lepton pair production in this kinematic region is dominated by the $\ln^2(Q^2/q_T^2)$-type of leading logarithmic contribution and can be calculated by using the Collins-Soper-Sterman (CSS) resummation formalism in the $b$-space, the Fourier transform of the $q_T$ space \cite{Collins:1984kg}.  Although the Fourier transformation from $b$-space calculation to the $q_T$-space could be sensitive to the non-perturbative physics at large $b$, the resummation formalism does have an excellent predictive power if the Fourier transform of the $b$-space distribution is dominated by the small $b$ region where the perturbative QCD calculation is effectively done in collinear factorization and reliable \cite{Qiu:2000ga}.  The CSS resummation formalism should work well for processes with a large hard scale $Q$ and large enough phase space for the gluon shower \cite{Berger:2004cc}.  
The formalism has been successful in interpreting the data 
on $W^\pm$ and $Z^0$ production in spin-averaged hadronic collisions \cite{Yuanetal}.  The generalization of the resummation formalism to the spin-dependent cross section with transverse-spin is nontrivial when the partonic hard part is not azimuthal symmetric, but, it should be a very interesting and important project for the future publication \cite{Boer:2001he,Idilbi:2004vb}.

\section*{Acknowledgments}

We thank C.~Gagliardi, T.~Kempel, J.~Lajoie, M.~Liu, E.~Sichtermann, and F.~Wei for helpful discussions.
This work was supported in part by the U. S. Department of Energy 
under Grant No.~DE-FG02-87ER40371. Z.~K. is 
grateful to RIKEN, Brookhaven National Laboratory, 
and the U.S. Department of Energy 
(Contract No.~DE-AC02-98CH10886) for providing the facilities essential 
for the completion of his work. 


\end{document}